\documentclass[%
reprint,
showpacs,
preprintnumbers,
amsmath,amssymb,
aps,
pra,
floatfix,
showkeys
]{revtex4-1}	
\usepackage[final]{graphicx}
\usepackage{dcolumn}
\usepackage{bm}
\usepackage[pdftex,rgb]{xcolor}
\usepackage{amsmath}
\usepackage[pdfstartview=FitV,      
            	breaklinks=true,        
            bookmarksopen=true,     
            bookmarksnumbered=true,  
            colorlinks,
            linkcolor = black, 
            citecolor = black, 
            filecolor = black,
            urlcolor = blue
            ]{hyperref}

\newcommand{\I}{\mathrm{i}}
\newcommand{\E}{\mathrm{e}}
\newcommand{\ket}[1]{|#1\rangle}

\newcommand{\braket}[1]{\langle #1\rangle}
\newcommand{\set}[1]{\lbrace #1\rbrace }
\newcommand{\dotkp}{\bm{k\mspace{1.5mu}{\cdot}\mspace{1.5mu}p}}
\DeclareSymbolFont{mathsmall}{U}{psy}{m}{n}
\DeclareMathSymbol{\varin}{\mathrel}{mathsmall}{206}

\begin{document}
\title{
	{Ground State Resonant Two--Photon Transitions\\\vspace*{.15cm} in Wurtzite GaN/AlN Quantum Dots}
	}
\date{
	\today
	}	
\author{\surname{Stefan T.} Jagsch}
\email[Contact: ]{jagsch@tu-berlin.de}
\affiliation{
	Institut f\"ur Festk\"orperphysik, 
	Technische Universit\"at Berlin, 
	Hardenbergstr.\,36,
	D-10623 Berlin,
	Germany}	
\author{\surname{Ludwig A. Th.} Greif}
\affiliation{
	Institut f\"ur Festk\"orperphysik, 
	Technische Universit\"at Berlin, 
	Hardenbergstr.\,36,
	D-10623 Berlin,
	Germany}	
\author{\surname{Stephan} Reitzenstein}
\affiliation{
	Institut f\"ur Festk\"orperphysik, 
	Technische Universit\"at Berlin, 
	Hardenbergstr.\,36,
	D-10623 Berlin,
	Germany}
\author{\surname{Andrei} Schliwa}
\affiliation{
	Institut f\"ur Festk\"orperphysik, 
	Technische Universit\"at Berlin, 
	Hardenbergstr.\,36,
	D-10623 Berlin,
	Germany}
	
\begin{abstract}
Two--photon transition rates are investigated in resonance to the ground state in wurtzite GaN/AlN quantum dots. The ground state transition is two--photon allowed because of the electron--hole separation inherent to polar wurtzite III--nitride heterostructures. We show that this built--in parity breaking mechanism can allow deterministic triggering of single--photon emission via coherent two--photon excitation. Radiative lifetimes obtained for single--photon relaxation are in good agreement with available time--resolved micro--photoluminescence experiments, indicating the reliability of the employed computational framework based on 8--band $\dotkp$--wavefunctions. Two--photon singly--induced emission is explored in terms of possible cavity and non--degeneracy enhancement of two--photon processes.
\end{abstract}

\maketitle

\section{Introduction}
Facilitated by strong carrier confinement and large exciton binding energies, III--nitride quantum dots (QDs) can show single--photon emission at room temperature and beyond.
Together with mechanical and thermal robustness, as well as chemical inertness, this makes them excellent candidates for future quantum--optoelectronics \cite{Hol19}. 
When grown along the [0001] crystallographic axis, wurtzite III--nitride systems exhibit strong internal pyro-- and piezoelectric fields of the order of MV/cm, constituting a pronouced quantum--confined Stark effect \cite{Bre06}.
The resulting separation of electrons and holes, which have an increased probability density at the top and bottom of the QD, respectively, leads to a reduced carrier overlap and prolonged radiative lifetimes are observed with increasing QD height \cite{Bre06}.
The charge carrier separation also mitigates parity selection rules for optical transitions. 
Usually forbidden transitions, such as the two--photon ground state transition between the lowest conduction band (CB) state and the highest valence band (VB) state, become allowed. 
In contrast to earlier works \cite{Rei08, Lin10}, where an external electric field is applied to a QD to observe parity forbidden transitions, the wavefunction parity in III--nitride QDs is broken by the built--in polarization fields.
\begin{figure}[!ht]
	\centering
	\includegraphics[width=.48\textwidth]{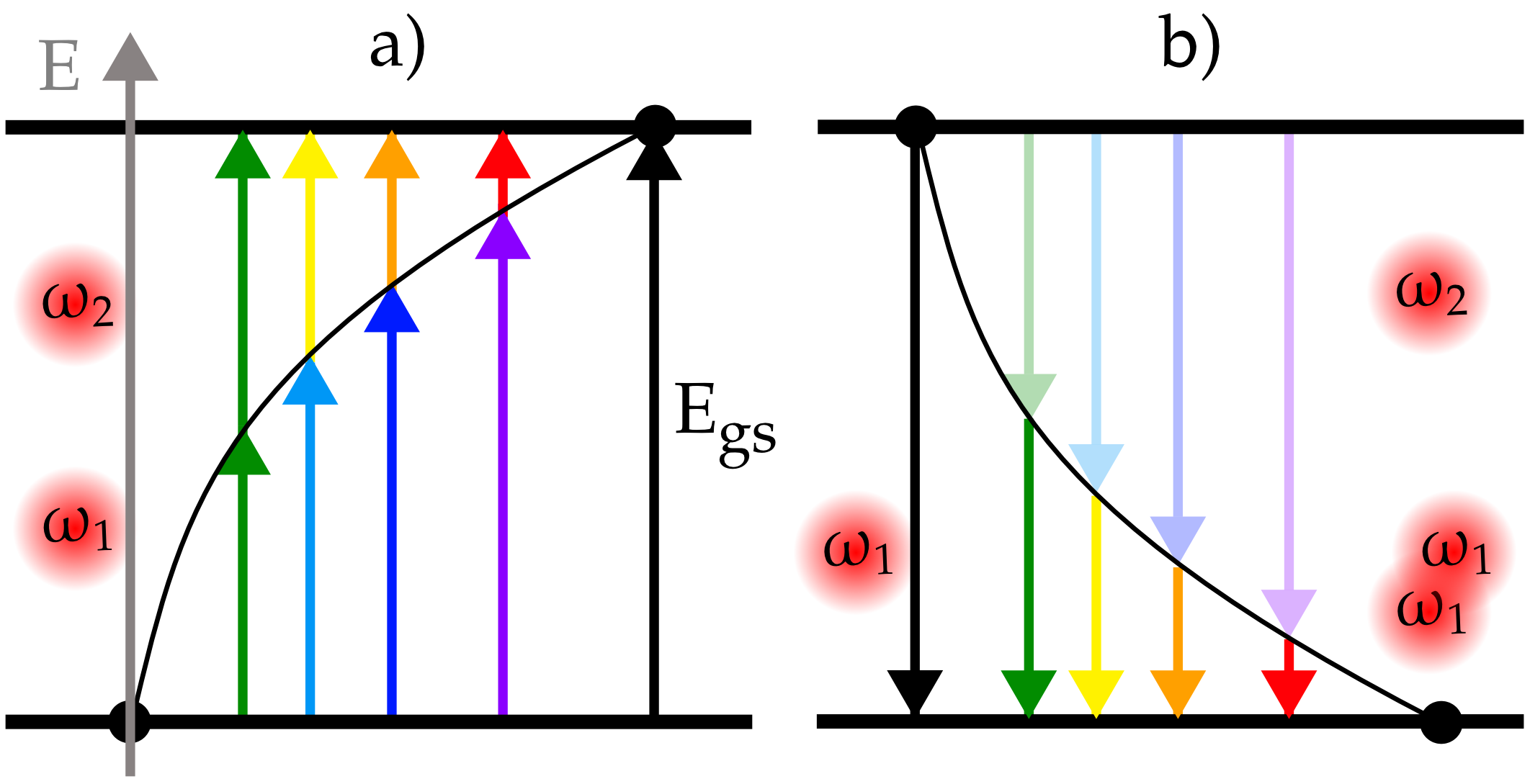}\\
	\caption{Schematic view of the two--photon processes investigated. a) Two--photon absorption. Two degenerate or non--degenerate input photons of frequency $\omega_{1}$ and $\omega_{2}$, with their combined energy tuned to the ground state transition energy $E_{gs}$, are absorbed to excite the QD. b) Two--photon singly--induced emission. A below band gap photon of frequency $\omega_{1}$ triggers the emission of an identical photon (second photon of frequency $\omega_{1}$) and a photon of the complementary frequency $\omega_{2}$, carrying the energy difference, from the excited QD.}
	\label{fig:tpatpie}
\end{figure}\\
State--of--the--art approaches to generating single photons on demand use almost exclusively either coherent (resonant) or incoherent single--photon excitation, both with their respective drawbacks. 
Incoherent excitation limits the indistinguishability of emitted single photons by phonon--assisted relaxation to the ground state, spectral jitter induced by trapped charges and populates the exciton bright and dark states with equal probability \cite{Tho16, Schm15}.
Coherent single--photon excitation overcomes these limiting issues, but instead requires an intricate suppression of excitation stray light in the detection path, at the cost of 50\,\% of the signal \cite{Som16, Din16}.
Herein, we employ the parity breaking mechanism inherent to polar wurtzite III--nitride QDs to trigger single photons coherently via two--photon excitation directly into the ground state. 
In principle, any pair of photons whose angular frequencies $\omega_{1}$ and $\omega_{2}$ satisfy the energy conservation requirement, $E_{gs}\,=\,\hbar(\omega_{1} +\omega_{2})$, can be employed for two--photon excitation (Fig.\,\ref{fig:tpatpie}a). Here, $E_{gs}$ is the ground state transition (band gap) energy. 
Generally, the excitation scheme could also be used to up--convert photons to the transition frequency when applying an appropriate complementary pump \cite{Fis11}.
Following an introduction of the theoretical framework, single--photon lifetimes are discussed and compared to available time--resolved micro--photoluminescence ($\mu$PL) data. 
Subsequently, we show that using two--photon resonant excitation, the ground state can be populated at a rate readily exceeding the radiative lifetime under realistic experimental conditions. 
Cavity and non--degeneracy enhancement of two--photon processes in QDs are discussed and  applied to singly--induced two--photon emission, where a below band gap photon is used to induce emission of an identical photon as well as a photon of complementary frequency (Fig.\,\ref{fig:tpatpie}b).

\section{Theoretical framework}
Within the dipole approximation light--matter interaction can be expressed in terms of the dipole operator $\bm{\mu}\,=\,-e\bm{r}$, with an interaction Hamiltonian given as $\bm{H}^{int}\,=\,-\bm{\mu\cdot E}(t)$, where $\bm{E}(t)$ is the electric field.
The invariance of transition rates between two states $\ket{f}$, $\ket{i}$ with respect to the velocity form based on the momentum operator $\bm{p}$ is discussed in Ref.\,\cite{Gol81} and expectation values transform according to \mbox{$\braket{f|\bm{p}|i}=\tfrac{\I m_{0}}{\hbar}(E_{f}-E_{i})\braket{f|\bm{r}|i}$}. Here $m_{0}$ denotes the electron mass and $E_{f,i}$ are the states' energies. From a perturbation ansatz to solving Schr\"odinger's equation, Fermi's Golden Rule 
\begin{align}
\label{eq:fgrlin}
	\Gamma_{\text{SP}}=\tau^{-1}=\frac{\omega_{gs}^{3}n}
	{3\hbar\pi c_{0}^{3}\varepsilon_{0}}|\bm{\mu}_{gs}|^{2}\,,
\end{align}
for the linear transition rate $\Gamma_{\text{SP}}$ is derived. 
Above, $\tau$ is the radiative lifetime, $n$ the refractive index, $c_{0}$ the vacuum speed of light, $\varepsilon_{0}$ the vacuum permittivity, $\omega_{gs}=E_{gs}/\hbar$ and \mbox{$\bm{\mu}_{gs}=\braket{h_{1}|\bm{\mu}|e_{1}}$} are QD ground state transition frequency and transition dipole, respectively, with $\ket{h_{1}}$ and $\ket{e_{1}}$ denoting the highest VB state and lowest CB state.
To arrive at Eqn.\,\ref{eq:fgrlin}, we considered the vacuum fluctuation field strength \mbox{$E^{(0)}_{\bm{k}}=\sqrt{\hbar \omega_{\bm{k}}/2\varepsilon_{0}n^{2}V}$} and the mode density of states (DOS) for an extended medium \mbox{$\rho_{dip}^{(0)}=\omega^{2}Vn^{3}/3\pi^{2}c_{0}^{3}$}, as seen by a dipole emitter. 
All transition dipoles herein are calculated using single--particle 8--band \mbox{$\dotkp$--wavefunctions} for the electron and hole states. 
The 8--band $\dotkp$--method for wurtzite crystal systems includes a realistic QD structure, strain and strain--induced piezoelectric fields as well as pyroelectricity, VB mixing and \mbox{CB--VB} coupling \cite{Win06, Sti00}. 
The theoretical framework employed to calculate the electronic states of the QDs is detailed in supplementary section A \cite{SM}. Material parameters are given in supplementary Tables I and II \cite{SM, Vur03}.
\begin{figure}[!ht]
	\centering
	\includegraphics[width=.48\textwidth]{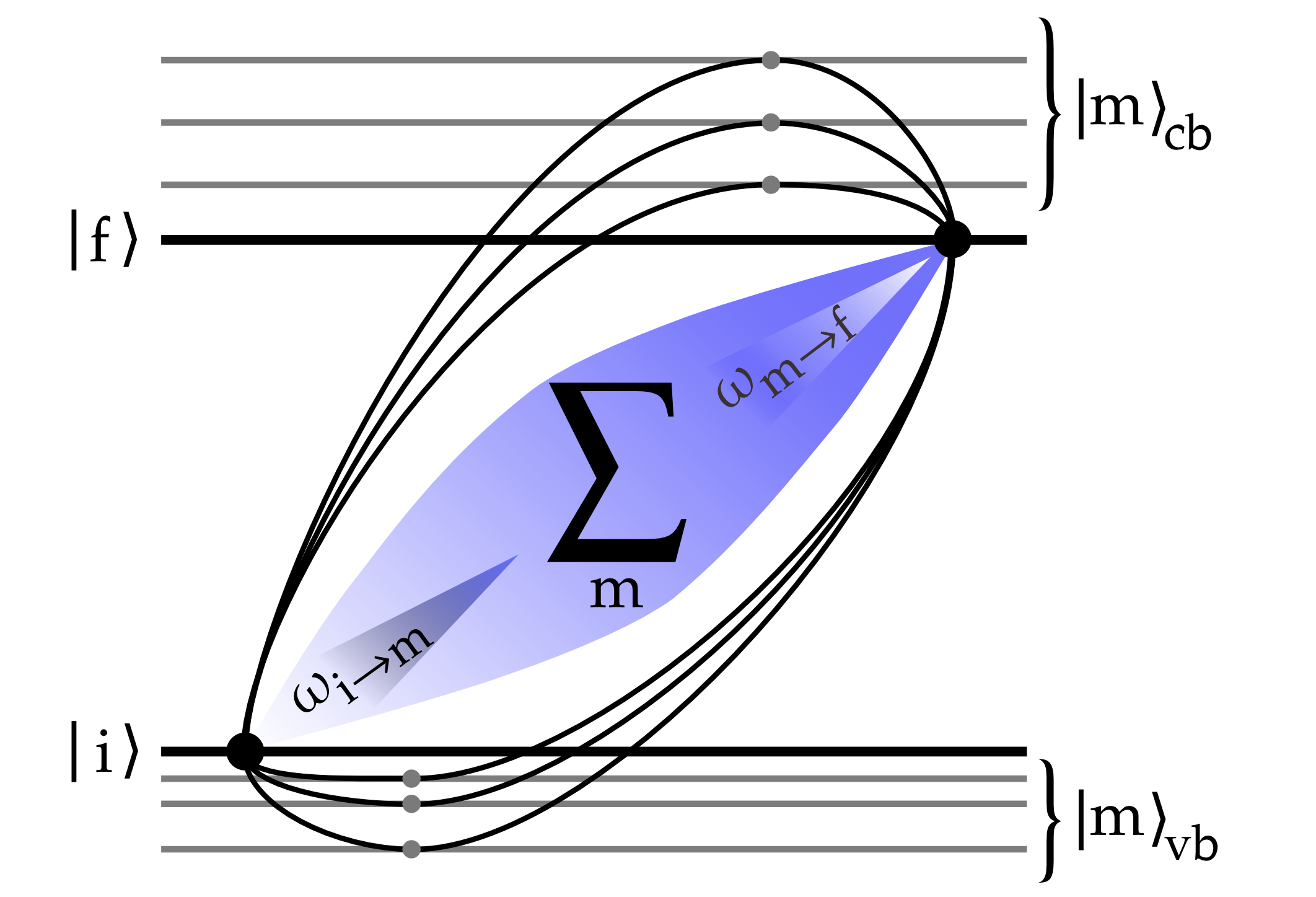}\\
	\caption{Schematic two--photon transition matrix element (shown for two--photon absorption). The second--order matrix element in Eqn.\,2 comprises the sum over all available intermediate states $\ket{m}$ in both VB and CB. Every intermediate state provides a channel $\ket{i}\rightarrow\ket{m}\rightarrow\ket{f}$, which are summed up to obtain the overall transition rate.}
	\label{fig:transitm}
\end{figure} 
As for any non--linear process, the transition matrix element for a two--photon process is a sum over transitions via a set of intermediate states $\set{\ket{m}}$, where each state $\ket{m}$ constitutes a channel to the overall transition from some initial state $\ket{i}$ to a final state $\ket{f}$ (schematic in Fig.\,\ref{fig:transitm}).
Within the rotating wave approximation, the second--order transition matrix element is of the form
\begin{align} 
\label{eq:matrixtpp}
\begin{aligned}
	M_{fi}=\sum\limits_{m} \Bigg\lbrace &
	\frac{(\bm{\mu}_{fm}\cdot\hat{\bm{\epsilon}}_{2})(\bm{\mu}_{mi}\cdot
	\hat{\bm{\epsilon}}_{1})}{\omega_{mi}\pm\omega_{1}}\\
	&+ \frac{(\bm{\mu}_{fm}\cdot\hat{\bm{\epsilon}}_{1})(\bm{\mu}_{mi}\cdot
	\hat{\bm{\epsilon}}_{2})}{\omega_{mi}\pm\omega_{2}}\Bigg\rbrace\,,
\end{aligned}
\end{align}
when describing two--photon processes in the general case of a dichromatic field $\bm{E}(t)=\frac{1}{2}(\bm{E}_{1}\E^{-\I\omega_{1}t} + \bm{E}_{2}\E^{-\I\omega_{2}t} + c.c.)$ \cite{Boy08}.
Above, $\omega_{mi}=(E_{m}-E_{i})/\hbar$ is the intermediate step frequency, $\hat{\bm{\epsilon}}_{1}$ and $\hat{\bm{\epsilon}}_{2}$ are the polarization unit vectors of the two field components with frequencies $\omega_{1}$ and $\omega_{2}$ and $-, +$ correspond to an absorption or emission process $\ket{i}\rightarrow\ket{f}$, respectively.
The calculation of the optical transition rates based on the $\dotkp$--wavefunctions is further described in supplementary section B \cite{SM}.
The set of intermediate states $\set{\ket{m}}$ comprises all real states of the system, irrespective of their occupation \cite{Ivc95}. Non--zero contributions to the transition arise if $\ket{m}$ is dipole--accessible from both $\ket{i}$ and $\ket{f}$.
The intermediate states $\ket{m}$ are referred to as virtual states, because intermediate transitions are not energy conserving and intermediate states cannot acquire an occupation themselves through the transition process.
Using the relation between intensity $I$ and the amplitude maximum of the corresponding real fields $E_{i}=\sqrt{2I_{i}/nc_{0}\varepsilon_{0}}$, the two--photon absorption (TPA) rate
\begin{align}
\label{eq:tpabulk}
	\Gamma_{\text{TPA}}=\frac{\pi}{2}\left[\frac{I_{1}}
	{\hbar^{2}nc_{0}\varepsilon_{0}}\right] \left[\frac{I_{2}}
	{\hbar^{2}nc_{0}\varepsilon_{0}}\right] |M_{fi}|^{2}\,   
	\delta_{fi}(\omega)
\end{align}
is obtained. We approximate the material's DOS at the transition frequency $\omega_{fi}=E_{fi}/\hbar$, i.\,e.\ the lineshape function of the QD state, with a delta function $\delta_{fi}(\omega)=\delta(\omega-|\omega_{fi}|)$, evaluated at $\omega=\omega_{1}+\omega_{2}$. \mbox{Equation \ref{eq:tpabulk}} is equivalent to the two--photon doubly--induced emission rate, which is the inverse process to TPA. 
Applying a pump field at a frequency $\omega_{1}<|\omega_{fi}|$ will induce emission at $\omega_{1}$ and the complementary frequency $\omega_{2}=|\omega_{fi}|-\omega_{1}$. From a quantum electrodynamic point of view, the emission probability is proportional to the mode occupation plus one, where the latter is the vacuum field fluctuation contribution. 
Therefore, two--photon emission is proportional to the simultaneous occupation of the two participating modes and the application of a field at one of the frequencies increases the rate \cite{Hay08}. Similar to Eqn.\,\ref{eq:tpabulk}, the two--photon singly--induced emission (TPIE) rate is given as
\begin{align}
\label{eq:tpiebulk}
	\Gamma_{\text{TPIE}}&=\frac{\pi}{2}\left[\frac{I_{1}}
	{\hbar^{2}nc_{0}\varepsilon_{0}}\right]
	\left[
	\frac{\omega_{2}^{3}n}{3\hbar\pi^{2}c^{3}_{0}\varepsilon_{0}}
	\right] |M_{fi}|^{2}\,  
	\delta_{fi}(\omega)\,.
\end{align}
The first bracket in Eqn.\,\ref{eq:tpiebulk} describes the inducing component at $\omega_{1}$ and the second bracket the induced component, containing mode DOS factor and vacuum fluctuation field strength at the complementary frequency.
Even in the absence of any inducing field, two--photon spontaneous emission (TPSE) can occur as a vanishingly low--signal, broadband background with contributions at any pair of frequencies satisfying energy conservation. Integrating over all contributions, the TPSE rate
\begin{align}
\label{eq:tpsebulk}
	\Gamma_{\text{TPSE}}=\frac{\pi}{4}\int\displaylimits_{0}^{|\omega_{fi}|}\left[
	\frac{\omega_{1}^{3}n}{3\hbar\pi^{2}c^{3}_{0}\varepsilon_{0}}
	\right] 
	\left[
	\frac{(|\omega_{fi}|-\omega_{1})^{3}n}{3\hbar\pi^{2}c^{3}_{0}\varepsilon_{0}}
	\right] |M_{fi}|^{2}\,\mathrm{d}\omega_{1}\,
\end{align}
is obtained. The polarization dependence of the matrix element in Eqn.\,\ref{eq:tpsebulk} was integrated out, yielding the mode DOS factors $\rho_{dip}^{(0)}$ that dictate the shape of the TPSE background \cite{Gol81, Hay08}.
Altering the mode DOS by an optical microcavity would subsequently lead to a spectral redistribution of the TPSE background with increased contributions at the cavity resonance and associated complementary frequency. In contrast to single--photon Purcell enhancement, both an increased mode DOS at resonance and possible off--resonant suppression need to be considered.\\
Unlike in higher dimensional structures, where dominant contributions to two--photon processes involve self--transitions, where $\ket{i}$ and $\ket{f}$ themselves can function as intermediate states \cite{Fis11, Whe84}, because of a finite photon momentum \cite{Bra64}, the discrete nature of the QD states forbids such contributions.
Generally, self--transition contributions are expected to be negligible close to the $\Gamma$--point, i.e. for direct semiconductors such as GaN/AlN \cite{Whe84}. 
Thus, in our theoretical approach the calculation of the transition rates cannot be simplified to self--transitions. 
Using the \mbox{$\dotkp$--framework} to calculate the individual transition dipoles also allows us to consider interference effects between the transition channels caused by the different orientations of the transition dipoles, as well as their energetic position with respect to the transition. 
Assuming a ground state two--photon transition, it is $\omega_{mi}<0,\,\forall\,\ket{m}\varin$ VB and $\omega_{mi}>\omega_{fm}>0,\,\forall\,\ket{m}\varin$ CB. Consequently, the denominator in Eqn.\,\ref{eq:matrixtpp} is negative for $\ket{m}\varin$ VB and positive for $\ket{m}\varin$ CB. 
Thus, independent of the process (emission or absorption), contributions from VB and CB intermediate states are phase shifted by $\pi$ \cite{Che11}.
Induced processes additionally maintain a polarization dependent channel interference through the transition dipole orientations which can yield both signs in the numerator for both types of intermediate states. A possible route to enhancing two--photon processes is to align the transition dipoles for channels with major contributions to the transition matrix element \cite{Cro02}.

\section{Results}
\begin{figure}[!ht]
	\centering
	\includegraphics[width=.48\textwidth]{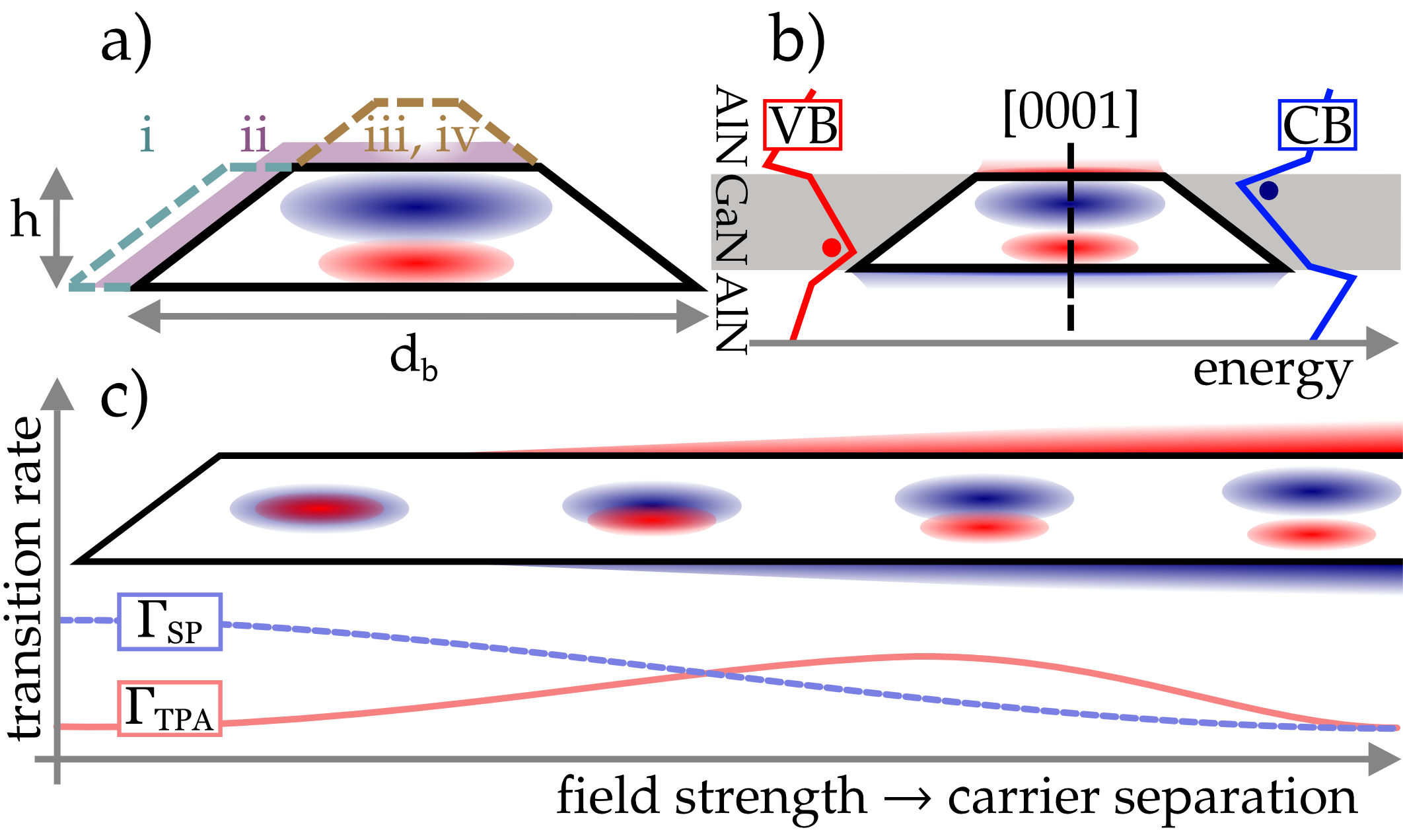}\\
	\caption{Qualitative sketches of the QD series, the field--induced carrier separation and the resulting influence on single--photon and two--photon transition rates. a) Four series of QDs are investigated, where either height (i), vertical aspect ratio $AR_{v}$ (ii) or base diameter $d_{b}$ (iii,\,iv) is kept constant. b) Schematic VB and CB energy structure as a result of polarization charges at the material interfaces. Electron (blue) and hole (red) probability density are vertically shifted along the [0001] direction. c) Exemplary influence of the carrier separation on both single--photon as well as two--photon transition rates.}
	\label{fig:qdseries}
\end{figure}

Radiative lifetime and two--photon transition rates are calculated for the QD ground state of a number of GaN/AlN QDs varying in height from 1 to 10 monolayers (MLs) and vertical aspect ratio $\mathrm{AR}_{v}$ (height $h$ to base diameter $d_{b}$) from 0.03 to 0.3, comprising values reported in literature \cite{Kak03, Bre06, Sim08, Gog03}.
The GaN QDs are modelled as hexagonal truncated pyramids with a side--wall angle of $30^{\circ}$ \cite{Kak06}, placed on a 1--ML--thick wetting layer and embedded in an AlN matrix. In the following, the transition rates are shown for four series of QDs where either base diameter, height or vertical aspect ratio is kept constant (Fig.\,\ref{fig:qdseries}a).
For a constant height of 5 MLs the base diameter is varied between 4 and 20\,nm in steps of 2\,nm (series i). For a constant vertical aspect ratio of $AR_{v}=0.2$, QDs with a height of 3, 5, 7 and 9 MLs are considered (series ii), in agreement with time--resolved $\mu$PL experiments in Ref.\,\cite{Kak03}. Finally, for two constant base diameters $d_{b}=12\,\mathrm{nm}$ and $16\,\mathrm{nm}$, the QD height is increased from 1 to 10 MLs in steps of 1 ML (series iii,\,iv).
\subsection{Single--photon radiative lifetime}
\begin{figure}[!ht]
	\centering
	\includegraphics[width=.48\textwidth]{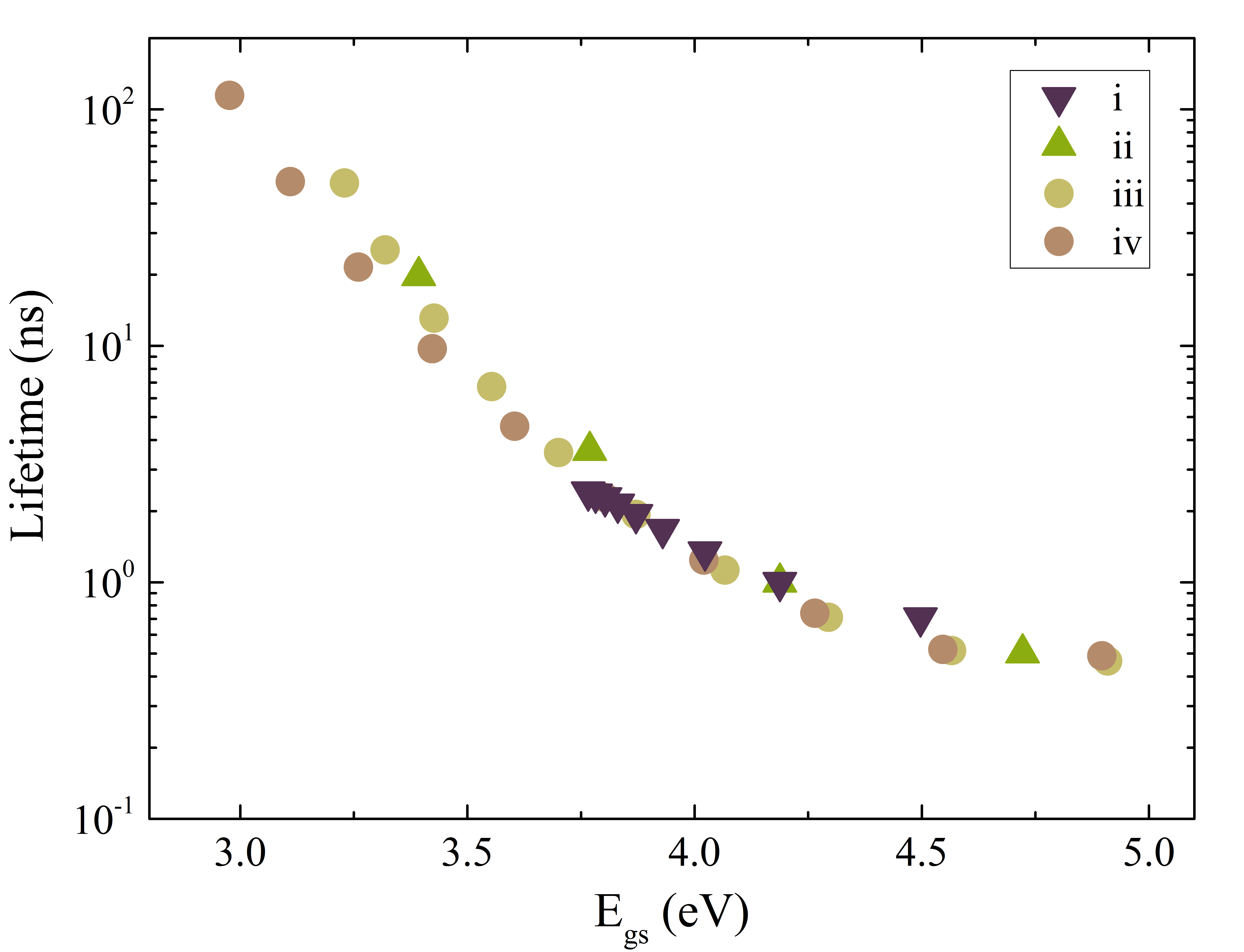}\\
	\caption{Single--photon lifetime in ns for the QD series i-v (Fig.\,\ref{fig:qdseries}a) as a function of the transition energy $E_{gs}$. A mono--exponential decrease in lifetime is observed up to a transition energy of about 4\,eV, before converging in the sub--ns regime. The data show only minor spread between the QD series.}
	\label{fig:opse}
\end{figure}

The single--photon radiative lifetime is calculated according to Eqn.\,\ref{eq:fgrlin} and shown in Fig.\,\ref{fig:opse} as a function of the ground state transition energy. 
A decrease in lifetime with increasing transition energy is observed, exhibiting little vertical spread caused by the different aspect ratios in the QD series. 
Generally, the lifetime is dominated by the height of the QD through the field--induced electron--hole separation and the resulting change in oscillator strength, proportional to the overlap integral (Fig.\,\ref{fig:qdseries}b,\,c). 
For the individual series, the lifetime decreases with increasing vertical and horizontal confinement, ultimately converging in the sub--ns regime \cite{Kak03}.
The influence of the horizontal confinement on the lifetime is more pronounced for higher QDs as a consequence of the side--wall angle and the associated shape anisotropy. 
The horizontal confinement for the electrons increases with the QD height as well as a reduced base diameter, with noticeable influence below a top diameter of $\sim$\,8\,nm.
Our results are in good agreement with time--resolved $\mu$PL measurements reported in Refs.\,\cite{Bre06, Kak03} in both order of magnitude as well as overall trend. See supplementary section C for a direct comparison \cite{SM}.
The agreement between experimentally obtained lifetimes and our results shows the reliability of the employed computational framework based on 8--band $\dotkp$--wavefunctions and justifies using single--particle states to calculate the transition rates.

\subsection{Two-photon excitation of the ground state}
\begin{figure}[!ht]
	\centering
	\includegraphics[width=.48\textwidth]{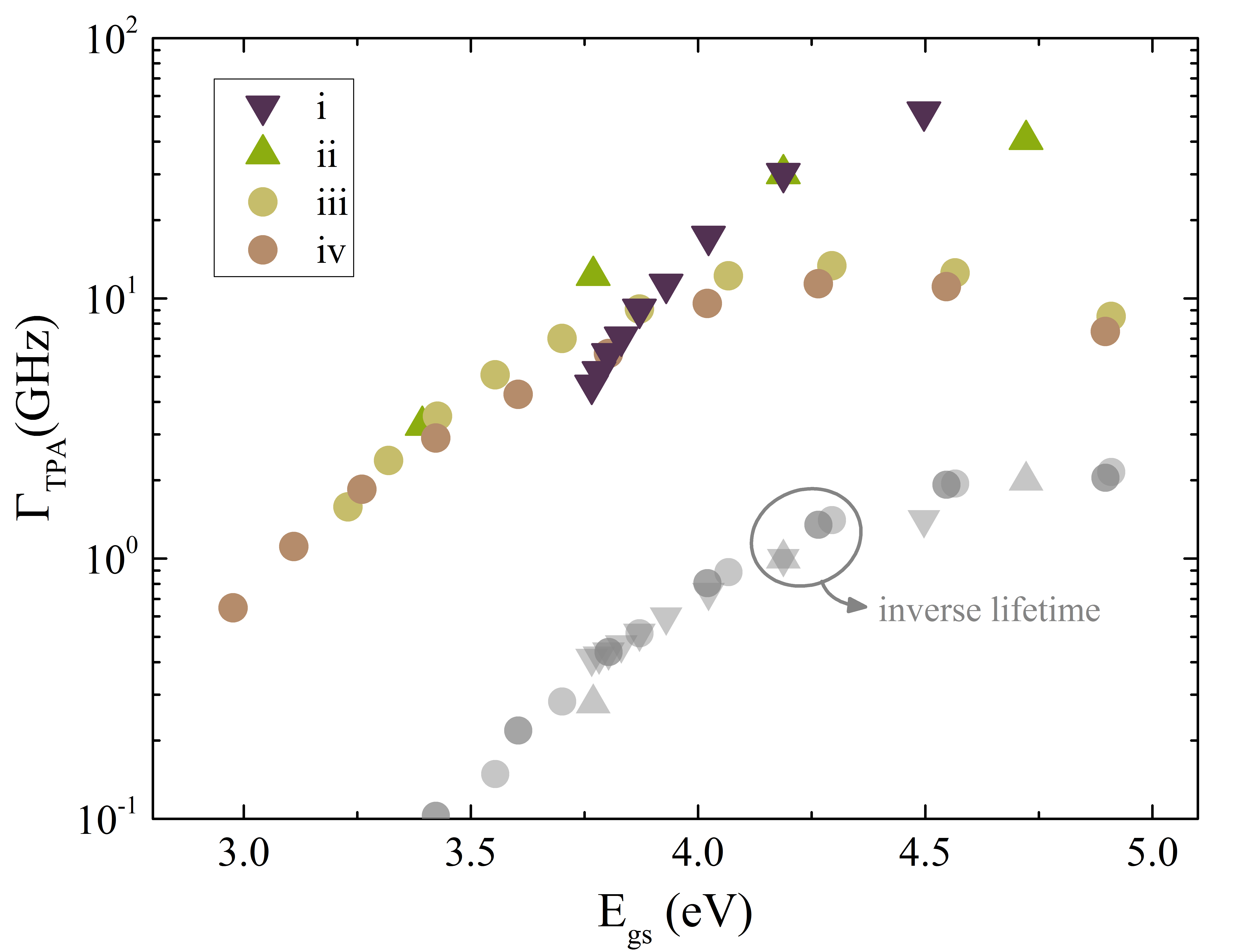}\\
	\caption{Two-photon absorption rate in GHz as a function of the transition energy $E_{gs}$ for the different QD series. Except for the constant height series (i), the series show a saturation (or eventual decrease) towards higher energies, indicating an optimal separation of electron and hole. The single--photon emission rate (inverse lifetime from Fig.\,\ref{fig:opse}) is added in greyscale for comparison.}
	\label{fig:tpa}
\end{figure}

The TPA rate is investigated (cf. Eqn.\,\ref{eq:tpabulk}), assuming an incident excitation power density of $I_{1}=I_{2}=1\mathrm{kW}/\mathrm{cm}^{2}$ with an incoupling efficiency of 5\%. Figure \ref{fig:tpa} shows the transition rate for degenerate input photons with their energy tuned to half of the ground state transition energy.
For the chosen parameters, the TPA rate already exceeds the single--photon emission rate (inverse single--photon lifetime from Fig.\,\ref{fig:opse}; shown in greyscale for comparison), suggesting that TPA can be used to resonantly pump the ground state transition. In experiment, the required pump rate additionally depends on the impact of non--radiative loss channels, which are quenched at cryogenic temperatures with the exception of material quality related carrier loss and Auger recombination. 
The two--photon absorption rate shows a saturating behavior with increasing transition energy and eventually starts to decrease again as a function of the QD height \mbox{(series iii \& iv)}. This behaviour indicates an optimal separation of electron and hole in the competition with parity selection suppression of TPA into the ground state (Fig.\,\ref{fig:qdseries}c) \cite{Rei08, Lin10}.
On the one hand, if the separation is too large (high QDs), the wavefunction overlap of ground state electron and hole with the intermediate states vanishes and with it their individual contribution to the TPA. 
On the other hand, if the QDs become very small, electron and hole ground state wavefunctions overlap increasingly and the two--photon transition becomes again parity forbidden. In comparison to the single--photon lifetime, the data show a larger relative spread between the QD series. 
The rate variation in the constant height series (i) suggests an increased influence of the vertical aspect ratio $AR_{v}$ on the transition rate. The intermediate states' envelope functions have increased probability densities away from the in--plane QD center and are consequently more sensitive to the lateral confinement. See supplementary Fig.\,1 for a visualization of the electron and hole states \cite{SM}. 
Hence, the overlap with the electron and hole ground state increases with reducing diameter. 
As a guideline for sample growth, QDs with a vertical aspect ratio of $AR_{v}\approx 0.2$ and emission energies of $\approx 4.5$\,eV (similar to the QDs realized in Ref.\,\citep{Kak03}) are the most promising candidates for an experimental investigation.
In the following, we discuss possible approaches to increasing two--photon processes in QDs, which will be necessary to achieve appreciable rates for singly--induced two--photon emission or to further enhance TPA.

\subsection{Non-degenerate two--photon processes and cavity enhancement}
From the two--photon transition matrix element, for a given pair of frequencies $\omega_{1}$ and $\omega_{2}$, cf. Eqn.\,\ref{eq:matrixtpp}, the contribution of any particular channel $\ket{m}$ to the transition $\ket{i}\rightarrow\ket{f}$ will depend upon several parameters. 
Those are energetic position $E_{m}$ with respect to the transition, magnitude of the transition dipoles and their orientation with respect to the field polarization(s).
However, the transition matrix element also strongly depends on the choice of input frequencies. For example, TPA shows a resonance enhancement for increasing non--degeneracy of the participating frequencies with an increasing transition rate until the linear absorption edge for the high energy component is reached. Here, the difference in photon energy can readily exceed a ratio of 1:10 for TPA \cite{Cir11, Fis11}.
Another benefit of using non--degenerate input frequencies is that multi--photon absorption into energetically higher states can be avoided by using the lower frequency beam as a stronger pump (or alternatively by using circularly polarized light). 
The same consideration applies to TPIE, where further enhancement can be expected if the inducing frequency $\omega_{1}$ (cf. Eqn.\,\ref{eq:tpiebulk}) is chosen low, because of the rate scaling $\propto (|\omega_{fi}|-\omega_{1})^{3}$.
We observe an enhancement of up to 2 orders of magnitude in case of TPA and even 3 orders of magnitude in case of TPIE for a frequency ratio of 1:5.\\
Alternatively, the two--photon rates can be enhanced by placing the QD in a suitable microcavity, thereby increasing the transition rate through cavity Purcell enhancement and an increased circulating intensity at resonance. 
Along the proposal in Ref.\,\cite{Lin10}, we assume a double--mode photonic crystal cavity with suitable resonances at both frequencies involved in the two--photon process. For the TPA rate an enhancement of up to
\begin{align}
\label{eq:tpacav}
	\frac{\Gamma^{cav}_{TPA}}{\Gamma_{TPA}}&=\left[\frac{\eta_{1}A_{cav}Q_{1}c}{\omega_{c1}V_{m1}n}\right]\left[\frac{\eta_{2}A_{cav}Q_{2}c}{\omega_{c2}V_{m2}n}\right]
\end{align}
is expected due to an increased circulating intensity at resonance $I\,\rightarrow\,I_{cav}=cQP_{in}/\omega_{c}nV_{m}$, where $Q$ is the quality factor, $V_{m}\approx(\lambda/n)^{3}$ the mode volume for a typical photonic crystal cavity and $P_{in}=\eta IA_{cav}$ with incoupling efficiency $\eta$ and illuminated cavity area $A_{cav}$. Effectively, both components rescale $\propto Q_{i}/\lambda_{i}^{2}$ when in resonance with the cavity ($\omega_{i}=\omega_{ci}$). 
The light--matter coupling strength for TPA can be approximated using an effective Rabi rate $\Omega_{TPA}=\sqrt{2\Gamma_{TPA}/\pi}$, which would indicate a transition into the strong coupling regime for values exceeding the dissipative loss rates in the system \cite{Lin10}. Analogously, the TPIE rate is expected to increase according to
\begin{align}
\label{eq:tpiecav}
	\frac{\Gamma_{TPIE}^{cav}}{\Gamma_{TPIE}}=\left[\frac{\eta_{1}A_{cav}Q_{1}c}{\omega_{c1}V_{m1}n}\right]\left[\frac{3Q_{2}\phi_{2}(\omega_{2})}{4\pi^{2}}\right],
\end{align}
benefiting from an increased energy density inside the cavity, as well as an increased mode DOS at the induced frequency. To arrive at Eqn.\,\ref{eq:tpiecav}, we approximated the cavity mode DOS as a normalized Lorentzian $\rho^{i}_{cav}(\omega)=2Q_{i}/\pi\omega_{ci}\cdot\phi_{i}(\omega)$ centred at $\omega_{i}$.

\subsection{Singly-induced two-photon emission}
Lastly, we examine the feasibility of singly--induced two--photon emission using a (far) below band gap pump beam. Here, we assume an excitation density of $I_{1}=10\,\mathrm{kW/cm^{2}}$. Since the pump beam is not absorbed in first order (further reduced for high non--degeneracy), a high input power density will not necessarily damage the material.
\begin{figure}[!ht]
	\centering
	\includegraphics[width=.48\textwidth]{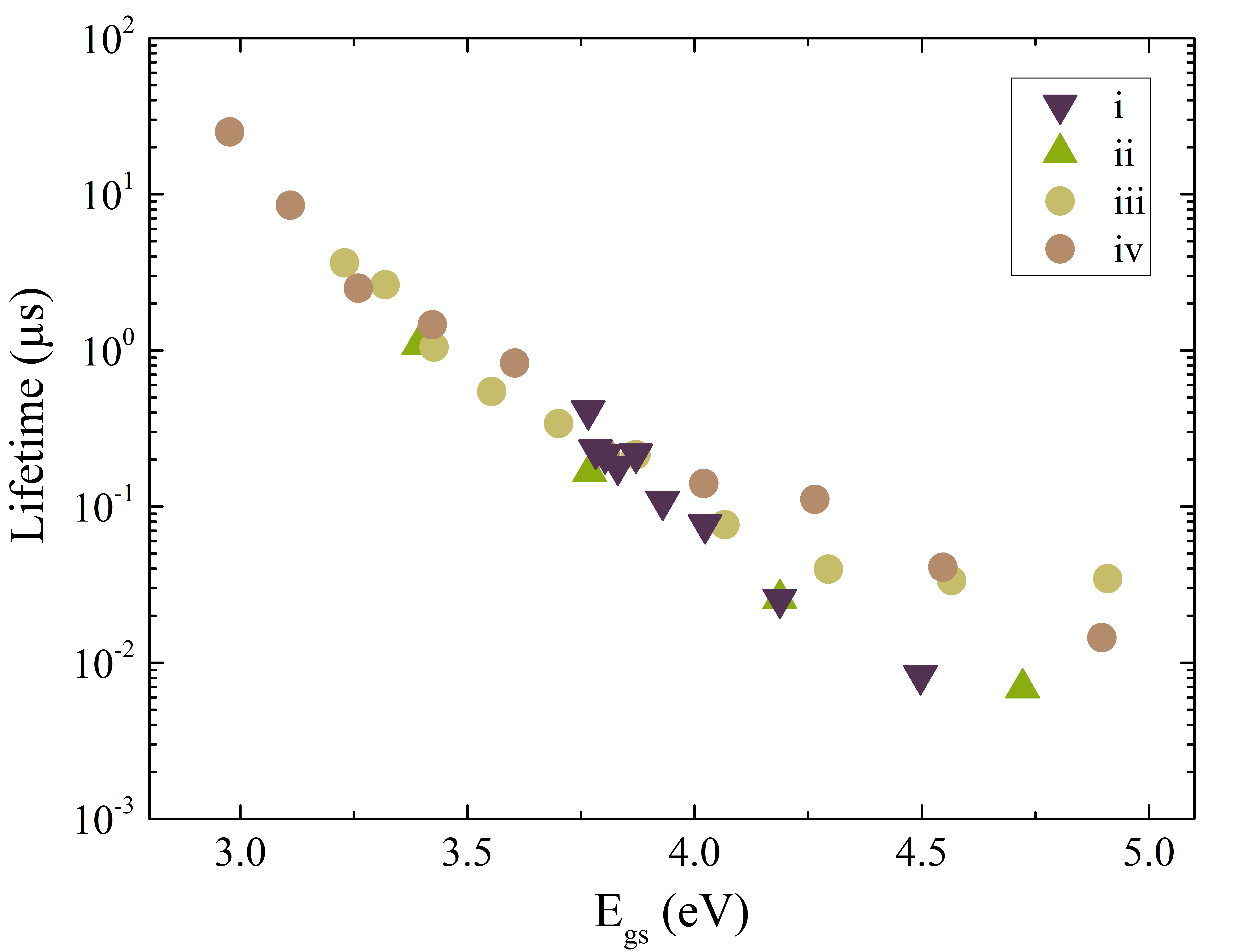}\\
	\caption{Two--photon singly--induced lifetime in $\mu$s as a function of the transition energy for a pump photon energy of 400\,meV. Slight variations between QDs are observed within the series, as a consequence of resonance enhancement if $\omega_{mi}\approx \omega_{1}$ for individual QDs.}
	\label{fig:tpie}
\end{figure}
Quality factors and mode volumes are chosen $Q_{1}=10^{2}$, $Q_{2}=10^{4}$ and $V_{m1}=V_{m2}=0.75(\lambda_{if}/n)^{3}$, respectively \cite{Lin10}. A possible mismatch between emitter and cavity mode field maximum was taken into account by further reducing the excitation efficiency by one order of magnitude.
In order for the TPIE signal to be spectrally separated in experiment, the pump frequency has to be detuned from half of the transition energy. 
In Fig.\,\ref{fig:tpie} we assume a pump photon energy of 400\,meV, which can be realized using a Ti:Sa OPO--system. 
The data shows an overall reduction in lifetime with increasing transition energy. 
However, the lifetime across the series is no longer smooth but instead exhibits minor increases or decreases from one data point to the next. The jumps are traced back to intermediate channels for which $\omega_{mi}\approx \omega_{1}$, resulting in a resonance enhancement for some QDs in the different series. 
The lifetimes reach values between 7\,ns and 25$\,\mu$s. Obviously, the lifetimes depend sensitively on the input parameters, such as $Q_{i}$, $V_{mi}$ and $\eta$ and whether or not the lifetime is short enough for the signal to be observable in experiment boils down to two main factors. 
Firstly, whether the TPIE complementary signal is spectrally separated from the single--photon transition energy and possible background luminescence, so that a low countrate can be compensated for by longer integration times. 
Secondly, when utilizing a microcavity to enhance TPIE, the single--photon transition energy can additionally be cavity suppressed. In this way, the competition between both processes is shifted in favour of the desired two--photon process for one combination of frequencies. 
The processing of appropriate samples, however, seems prohibitive when thinking beyond proof of principle experiments or of possible applications such as frequency up--conversion.

\section{Conclusion}
We discuss the mitigation of parity selection rules for optical transitions in wurtzite GaN/AlN QDs. 
The crucial aspect is the electron--hole separation inherent to polar wurtzite III--nitride heterostructures. 
As a result, the ground state transition is two--photon allowed.
We propose to employ this built--in parity breaking mechanism to trigger single--photon emission coherently via two--photon excitation.
Enhancement of two--photon processes for non--degenerate input frequencies and QDs in optical microcavities is discussed and could make this approach feasible. 
As an experimental guideline for two--photon processes, cavity and non--degeneracy enhancement are promising routes for increasing the transition rate and may allow the observation of singly--induced two--photon emission from (single) QDs.

\acknowledgements{The research leading to these results has received funding from the German Research Foundation (DFG) in the framework of project No. RE2974/8-1.}

 
\end{document}